\documentclass[twocolumn,showpacs,preprintnumbers,amsmath,amssymb]{revtex4}

\usepackage{graphicx}
\usepackage{dcolumn}
\usepackage{bm}

\newcommand{\be}{\begin{equation}}
\newcommand{\ee}{\end{equation}}
\newcommand{\bdis}{\begin{displaymath}}
\newcommand{\edis}{\end{displaymath}}
\newcommand{\pd}{\partial}

\begin{document}

\title{Study of the Coleman - de Luccia instanton of the second order}

\author{Michal Demetrian}\email{demetrian@fmph.uniba.sk}
\affiliation{Comenius University \\
Mlynska Dolina F2, 842 48, Bratislava, Slovak Republic }

\date{\today}

\begin{abstract}
We study the second order Coleman - de Luccia instanton which appears as the curvature of the effective potential
reaches a sufficiently large value.
We show how one can find the approximative formula for this instanton by perturbative
expansion in the case when the second derivative of the effective potential divided by the Hubble
parameter squared is close to $-10$, and we perform a numerical study of this instanton in the case of
quasi-exponential potential.
\end{abstract}

\maketitle

\section{Introduction}
The theory of false vacuum decay in de Sitter universe via formation of rapidly expanding bubbles
was introduced by Coleman and de Luccia \cite{cdl} and it has been studied intensively as an
ingredient of old inflation \cite{guth} and of open inflation \cite{{oi1},{oi2}}. This mode of vacuum
decay is described by the Coleman - de Luccia (CdL) instanton - the nontrivial finite-action O(4) solution
to the Euclidean equations for coupled scalar and gravitational fields,
with an effective potential containing a true vacuum (at which it has
the global minimum equal to zero) and a false vacuum (at which it has a
local minimum) separated from the true vacuum by a finite potential
barrier. It is given by two functions $\Phi (\tau)$ and $a (\tau)$, where
$\Phi$ is the scalar field, $a$ is the radius of the 3-spheres of homogeneity
determined from the circumference and $\tau$ is the radius of the 3-spheres
of homogeneity measured from the centre.
Denote the effective potential by $V$. Functions $\Phi$ and $a$ obey the equations
\be\label{ee}
\ddot{\Phi} + \frac{3\dot{a}}{a}=V',\
\ddot{a}=-\frac{8\pi}{3}\left(\dot{\Phi}^2+V\right)a,
\ee
where the overdot denotes differentiation with respect to $\tau$ and the prime
differentiation with respect to $\Phi$. The boundary conditions ensuring finiteness of the (Euclidean) action
\bdis
S_E= \int{\rm d}^4x\ \sqrt{|g_E|}\left[ \frac{1}{2}g_E(\pd\Phi,\pd\Phi)+V(\Phi)-
\frac{R\left( g_E\right)}{16\pi}\right].
\edis
are
\be \label{eebc}
a(0)=\dot{\Phi}(0)=\dot{\Phi}(\tau_f)=0 \ \mbox{and} \ \dot{a}(0)=1,
\ee
where $\tau_f>0$ is defined by the equation $a(\tau_f)=0$. \\

Being motivated by the arguments from the papers \cite{{hm},{linde},{js},{gl}} the authors of the paper
\cite{vladoaja} have proved both
a necessary and a sufficient condition for the existence of the CdL instanton.
Before we formulate them let us introduce useful notations which will be used throughout the paper: by
$V_M=V(\Phi_M)$ we denote the local maximum of $V$
(the top of the barrier); and by $H(\Phi)$ we mean the value of the Hubble parameter corresponding to the energy
density equal to $V(\Phi)$, $H(\Phi)=\sqrt{8\pi V(\Phi)/3}$. The conditions read: \\

{\it If the inequality $-V''_M/H_M^2>4$ holds then the CdL instanton exists. } \\

{\it If the CdL instanton exists then the inequality $-V''(\Phi)/H^2(\Phi)>4$ holds somewhere in the barrier. } \\

For a given theory (potential) the conditions leave a gray zone in the parameter space of the potential in which
they cannot decide whether the CdL instanton exists or not. It has been shown in \cite{vladoaja}
that in addition to the first order instanton, also
instantons of higher order exist if the curvature of the potential in its local maximum is sufficiently large.
Namely, any finite-action solution of Eqs. (\ref{ee}) can be characterized by how many times the scalar field
crosses $\Phi_M$. If the scalar field crosses $\Phi_M$ $l$ times, we shall call the solution
the $l$th order CdL instanton. From the analysis of the instantons localized close to the top of the barrier
\cite{vladoaja} it follows that \\

{\it If $-V_M''/H_M^2>l(l+3)$ then the potential admits the $l$th order instanton.} \\

For odd $l$ there is, for given $V$, at most one instanton of $l$th order; for even $l$ we have two
$l$th order instantons. One of them starts and ends at $\Phi_i^L$ less than $\Phi_M$
and the other one starts and ends at $\Phi_i^R$ greater than $\Phi_M$.

\section{Analytic approach to the second order CdL instanton localized close to the top of the potential barrier}

The aim of this paragraph is to solve perturbatively Eqs. (\ref{ee}) in the case when
$-V_M''/H_M^2\approx 10$. Our task is to sketch how to get such a solution.
The detailed explanation
of this method (together with its application to the first order CdL instanton) will appear in
\cite{vladoaja2}.

It is useful to introduce the notations
\begin{eqnarray*}
& & y(x)\equiv\Phi(x)-\Phi_M=\sum k^n u_n(x) \\
& & a(x)=CH_M^{-1}\sum k^nv_n(x) \\
& & -\frac{V_M''}{H_M^2}=10+\sum k^n \Delta_n ,
\end{eqnarray*}
where we are using the dimensionless Euclidean time $x=H_M\tau$ that runs from $0$ to $\pi$, $C=8\pi/3$, and
$k$ is our parameter of perturbative expansion - i.e. the amplitude of $\Phi-\Phi_M$. We will continue our
calculations up to the order in which we will be able to decide what quantity seems to be interesting
or helpful if one would like to construct (numerically) the second order instanton in a given theory.
The functions $u_n$ and $v_n$ obey differential equations
\begin{eqnarray}
& & u_n''(x)+3\cot(x)u_n'(x)+10u_n(x)=\mathcal{U}_n(x), \nonumber \\
& & v_n''(x)+v_n(x)=\mathcal{V}_n(x)\sin(x),
\end{eqnarray}
where the
functions $\mathcal{U}_n$ and $\mathcal{V}_n$ can be computed order by order from the system of equations
\begin{eqnarray*}
& &
y''(x)+\frac{3a'(x)}{a(x)}y'(x)=\frac{V'(y)}{H_M^2}, \\
& &
a''(x)=-C\left(y'^2(x)+\frac{V(y)}{H_M^2}\right)a(x).
\end{eqnarray*}
Since $\mathcal{U}_1$ is identically equal to zero, $u_1$ is the nonsingular (at the end points) solution of
the equation
\bdis
u_1''(x)+3\cot(x)u_1'(x)+10u_1(x)=0
\edis
i.e.
\be \label{fciau1}
u_1(x)=\frac{1}{4}\left( 5\cos^2(x)-1\right).
\ee
To get $v_2$ one  has to find out that $\mathcal{V}_2=-(u_{1}')^{2}+5u_1^2$ and therefore $v_2$ obeys
\begin{eqnarray*}
& &
v_2''(x)+v_2(x)=\\
& &
\frac{5}{128}\left( -7\sin(x)+\frac{15}{2}\sin(3x)+\frac{45}{2}\sin(5x)\right)
\end{eqnarray*}
with the initial conditions $v_2(0)=v_2'(0)=0$. We straightforwardly obtain that
\begin{eqnarray} \label{fciav2}
& & v_2(x)= \frac{5}{256}\left(7x\cos(x)-\frac{15}{8}(\sin(3x)+\sin(5x))\right. \nonumber \\
& &
 +8\sin(x)\Big) .
\end{eqnarray}
We can proceed to $u_2$; in order to do this we need to know the function $\mathcal{U}_2$,
\bdis
\mathcal{U}_2(x)=-\Delta_1u_1(x)+\frac{1}{2}\eta u_1^2(x) ,
\edis
where $\eta=V_M'''/H_M^2$. The equation for $u_2$ reads
\begin{eqnarray*}
& & u_2''(x)+3\cot(x)u_2'(x)+10u_2(x)=-\frac{\Delta_1}{4}(5\cos^2(x)-1) \\
& & + \frac{\eta}{32}(5\cos^2(x)-1)^2 .
\end{eqnarray*}
The last equation can be transformed to an equation of hypergeometric type by changing the independent variable
$z=\cos(x)$. In this way one gets
\begin{eqnarray*}
& & (1-z^2)u_2''(z)-4zu_2'(z)+10u_2(z)= \\
& & -\frac{\Delta_1}{4}(5z^2-1)+\frac{\eta}{32}(5z^2-1)^2.
\end{eqnarray*}
We need a solution of this equation which is bounded (at the points $z=\pm 1$). Such a solution is given
by a polynomial; one can easily see that it must have the form $\alpha+\beta z^4$. Inserting the previous ansatz
in to the equation we get a system of three linear equations for $\alpha,\ \beta$ and $\Delta_1$. These equations
are parametrized by $\eta$ and their solution reads
\be \label{Delta1so}
\Delta_1=\frac{1}{6}\eta
\ee
and
\be \label{fciau2}
u_2(x)=\frac{\eta}{192}\left(\frac{7}{5}-\frac{25}{3}\cos^4(x)\right).
\ee
The "quantization rule" for $k$ follows from Eq. (\ref{Delta1so}),
\be \label{kso}
k=-\frac{6}{\eta}\left(\frac{V_M''}{H_M^2}+10\right).
\ee
It is also possible, within our perturbative approach,
to answer the question how the action of our second order CdL instanton differs from that
quantity of related Hawking - Moss instanton. In order to find this difference
one has to proceed to the third order calculations and
obtain the function $v_3$. However, we are not going to do this. Comparison of the two
actions is going to appear in \cite{vladoaja2}. \\

We can conclude that, in addition to the parameter
$V_M''/H_M^2+10$, our results depend crucially on the
value of the third derivative of the potential at $\Phi_M$. Moreover, we can see from Eq. (\ref{kso})
that these results do not hold
in the case $V_M'''=0$ and they require, for a given $V_M''/H_M^2+10$, a sufficiently large value of
$\eta$. To determine approximative behaviour of the second order CdL instanton in the case when $\eta$ is close to zero
one should perform extended analysis including the solution of the instanton's equations, at least,
up to the third order in $k$ (to determine $\Delta_2$).
Instead of doing this, we are going to investigate the properties of the
second order CdL instanton numerically, and we understand previously presented steps as a help in numerical
calculations. Namely, they showed us that the third derivative of the effective potential in its local maximum can be
regarded as an interesting quantity.

\section{Numerical study of the second order CdL instanton}

Let us consider, as an example, the theory with the potential
\be \label{qepot}
V(\Phi)=\left(V_0+\frac{1}{2}\Phi^2\right)e^{-\Phi/\Phi_0} ,
\ee
where $V_0$ and $\Phi_0$ are positive constants. The motivation for this potential can be found in \cite{barrow}.
The potential (\ref{qepot}) is also very convenient
for the investigation of the CdL instanton of the second order because it
allows us to preserve the barrier while $V_M'''$ is running from negative to positive values; quartic
potential has not this property. The barrier in the theory (\ref{qepot}) exists only if $V_0<1/2\Phi_0^2$,
and the values of $\Phi$ at which $V$ has the local maximum and local minimum respectively are
\bdis
\Phi_{m,M}=\Phi_0(1\mp \Delta) ,
\edis
where we have introduced the parametrization of the potential by $(\Phi_0,\Delta)$ with
\bdis
\Delta=\sqrt{1-2\frac{V_0}{\Phi_0^2}}, \quad \Delta\in[0,1].
\edis

\begin{figure}[h]
\includegraphics[width=9cm,height=6cm]{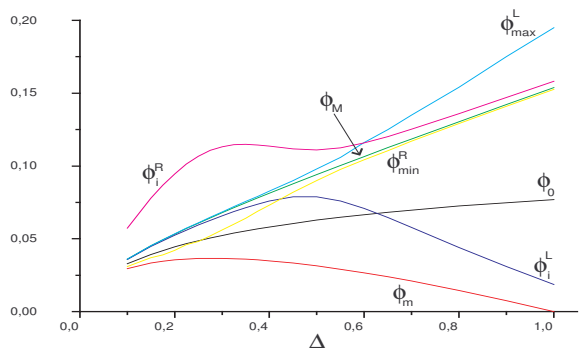}
\caption{Dependence of the second order CdL instantons in the theory (\ref{qepot}) on the parameter $\Delta$.
Description of the quantities introduced in this figure: $\Phi_i^L$ is the initial and final value
of $\Phi$ for the instanton starting left from $\Phi_M$, $\Phi^L_{max}$ is the maximal value of $\Phi$ for this
instanton; $\Phi_i^R$ is the initial and final value of $\Phi$ for the instanton starting right from $\Phi_M$ and
$\Phi^R_{min}$ is the minimal value of $\Phi$ for this instanton. We can see that the right-handed instanton gets
shrank (in the $\Phi$ direction) towards $\Phi_M$ as $\Delta$ approaches $1$ and the left-handed instanton is
the narrowest as $\Delta$ approaches $0$. The absolute value of the difference $\Phi_i^L-\Phi_M$ falls down as
$1/V_M'''$ for $\Delta\to 0^+$ and analogically $\Phi_i^R-\Phi_M$ falls down for $\Delta\to 1^-$. This means that
in the case $V_M'''<0$ ($V_M'''>0$)
the left (right)-handed instanton is the instanton that approaches the limit instanton as $-V_M''/H_M^2-10\to 0$.}
\label{f1}
\end{figure}

The quantities interesting for our goal are
\bdis
-\frac{V_M''}{H_M^2}=\frac{3\Delta}{8\pi\Phi_0^2(1+\Delta)}
\edis
and
\bdis
V_M'''=\frac{2\Delta-1}{\Phi_0^2}\exp(-1-\Delta).
\edis
We have put in our numerical computations $-V_M''/H_M^2-10=0.1$. We investigated parameters of the CdL instantons
of the second order as functions of the parameter $\Delta$ in its range of definition. Remind that there are
always two such instantons - see the introduction. The results are plotted on the fig. \ref{f1}.

\acknowledgments{The author thanks Vlado Balek for useful discussions. This paper was supported by the
grant VEGA 1/3042/06.}

\end{document}